\documentclass[conference]{IEEEtran}
\IEEEoverridecommandlockouts
\usepackage{cite}
\usepackage{amsmath,amssymb,amsfonts}
\usepackage{algorithmic}
\usepackage{graphicx}
\usepackage{textcomp}
\usepackage{xcolor}
\usepackage{booktabs}
\usepackage{threeparttable}
\def\BibTeX{{\rm B\kern-.05em{\sc i\kern-.025em b}\kern-.08em
    T\kern-.1667em\lower.7ex\hbox{E}\kern-.125emX}}

\begin{document}

\title{Evaluation of Look-ahead Economic Dispatch Using Reinforcement Learning    \\
\thanks{This work is supported by State Grid Corporation of China Project "Research on Key Technologies of Data-driven Based Auxiliary Decision Making of Look-ahead Power Dispatch".}
}

\author{
    \IEEEauthorblockN{Zekuan Yu}
    \IEEEauthorblockA{
    \textit{Department of Electrical Engineering} \\
    \textit{Tsinghua University} \\
    Beijing 100084, China
    }
    \and
    \IEEEauthorblockN{Guangchun Ruan}
    \IEEEauthorblockA{
    \textit{Department of Electrical Engineering} \\
    \textit{Tsinghua University} \\
    Beijing 100084, China
    }
    \and
    \IEEEauthorblockN{Xinyue Wang}
    \IEEEauthorblockA{
    \textit{Department of Electrical Engineering} \\
    \textit{Tsinghua University} \\
    Beijing 100084, China
    }
    \and
    \IEEEauthorblockN{Guanglun Zhang}
    \IEEEauthorblockA{
    \textit{Department of Electrical Engineering} \\
    \textit{Tsinghua University} \\
    Beijing 100084, China
    }
    \and
    \IEEEauthorblockN{Yiliu He}
    \IEEEauthorblockA{
    \textit{Department of Electrical Engineering} \\
    \textit{Tsinghua University} \\
    Beijing 100084, China
    }
    \and
    \IEEEauthorblockN{Haiwang Zhong}
    \IEEEauthorblockA{
    \textit{Department of Electrical Engineering} \\
    \textit{Sichuan Energy Internet Research Institute} \\
    \textit{Tsinghua University} \\
    Beijing 100084, China \\
    zhonghw@tsinghua.edu.cn
    }
}

\maketitle

\begin{abstract}
Modern power systems are experiencing a variety of challenges driven by renewable energy, which calls for developing novel dispatch methods such as reinforcement learning (RL). Evaluation of these methods as well as the RL agents are largely under explored. In this paper, we propose an evaluation approach to analyze the performance of RL agents in a look-ahead economic dispatch scheme. This approach is conducted by scanning multiple operational scenarios. In particular, a scenario generation method is developed to generate the network scenarios and demand scenarios for evaluation, and network structures are aggregated according to the change rates of power flow. Then several metrics are defined to evaluate the agents' performance from the perspective of economy and security. In the case study, we use a modified IEEE 30-bus system to illustrate the effectiveness of the proposed evaluation approach, and the simulation results reveal good and rapid adaptation to different scenarios. The comparison between different RL agents is also informative to offer advice for a better design of the learning strategies.
\end{abstract}


\section{Introduction}

The rapid decarbonization of modern power systems in recent decades presents a variety of operational challenges~\cite{papadis2020challenges}. These systems are undergoing a rapid growth by integrating more intermittent renewable energy sources, energy storages, and flexible loads. Here, the growing number and complexity of system components makes it significantly difficult to maintain the secure and cost-effective operation of power systems. Under this trend, traditional methods adapt and perform poorly because they are inflexible and only fit for typical working conditions. It is thus a urgent need in modern power systems to develop the next-generation dispatch methodologies.

Reinforcement learning (RL) provides an emerging and promising option to tackle the existing challenges~\cite{ruan2020review}. In an RL framework, an agent is trained to mimic the decision process of real system operators---they interact with a virtual environment of power systems and learn from the past decisions and rewards \cite{zhangDeepReinforcementLearning2020}. Making use of the hidden data pattern, a well-trained agent is capable of accommodating for multiple resources in a reasonable variety of scenarios. Many scholars have explored the viability of applying RL algorithms to power system dispatch. The authors of \cite{tieJointOptimizationDispatching2019} applied a deep deterministic policy gradient (DDPG) algorithm to the solution of a joint dispatch problem considering both traditional thermal units and renewable resources. In \cite{liuDistributedEconomicDispatch2018}, a cooperative RL algorithm was proposed to achieve efficient distributed economic dispatch in a microgrid with energy storage systems. An improved DDPG algorithm was developed in \cite{yangDynamicEnergyDispatch2021} to solve the dynamic economic dispatch problem for integrated energy systems. 

Despite the growing number of research on RL-based power system economic dispatch, the evaluation approach has not been fully studied. The authors of~\cite{maihemuti2022new} focused on the security evaluation issue for renewable-rich power systems using SWOT analysis. In~\cite{villanueva2011simulation}, a simulation approach was developed to improve the evaluation by taking the uncertain wind speed into consideration.
Often, performance evaluation consists of a novel step in economic dispatch, and then compares the algorithm with traditional baselines to validate its effectiveness \cite{nematiOptimizationUnitCommitment2018, sunSolvingPowerEconomic2014}. Some research works have been carried out on this topic. In \cite{8395546}, multiple data-driven algorithms for economic dispatch were evaluated from the aspect of optimality and computational complexity, and the advantage of these algorithms was discussed as well. In \cite{choiDataPerturbationBasedSensitivity2017}, the influence of data corruption on the look-ahead economic dispatch decisions was evaluated using sensitivity analysis, and a linear sensitivity matrix was derived to make a fast evaluation. 
To sum up, the evaluation approach for RL in power system economic dispatch remains under explored, especially concerning its unique ability to adapt to various scenarios. 

In this paper, we propose an evaluation approach for look-ahead economic dispatch to analyze the performance of RL agents faced with multiple scenarios. To be specific, the agent is designed for the N-1 contingency management, and the network scenarios for evaluation are selected from the N-1 cases by network clustering. Based on the agents' dispatch decisions, several metrics are defined to evaluate their performance. This evaluation approach will simulate with different parameters and inform the designs of learning strategies. The contributions of this paper are summarized as follows:

\begin{itemize}
    \item A network clustering method is proposed to cluster and aggregate the network structures before finalizing the network scenarios. The average change rates of power flow on critical transmission lines are utilized as features of the network structures for clustering, representing the consequences and influence of line outage.
    \item A group of evaluation metrics are developed to comprehensively analyze the agents' performance. The metrics for security are defined based on the limits of the constraints, while the metrics for economy are defined with respect to the baseline of dispatch decisions.
    \item The effectiveness of the proposed evaluation approach is validated by showing the adaptation ability of the past agents in various scenarios. In addition, the proposed approach can be utilized to offer suggestions for learning strategies.
\end{itemize}

The remainder of this paper is organized as follows. The scenario generation method based on network clustering is formulated in Section II. Section III introduces the evaluation metrics and the corresponding baseline generation approach. Case study based on a modified IEEE 30-bus system is presented in Section IV, and Section V concludes this paper.

\section{Scenario Generation based on Network Clustering}

In this section, multiple scenarios are generated for evaluation of the agent by combining network scenarios and demand scenarios. All the scenarios are based on the original conditions used for training the agent. The generation of demand scenarios is achieved by imposing a series of random variations upon the original demand level, while the generation of network scenarios is accomplished by aggregating network structures with similar features.

As the agent is designed for N-1 contingency, the potential network scenarios are limited to the N-1 cases, where only one transmission line is out of service. Compared with the N case where all transmission lines are functional, the power flow on the line with outage is redistributed to other lines, which results in the shift of power flow distribution, complying with the values of power transfer distribution factors \cite{tejada-arangoSecurityConstrainedUnit2018}. Such shift is utilized here as a feature of the N-1 case and its corresponding network for network clustering.

For an N-1 case with an outage in line $l$, the observation vector is defined as follows:
\begin{align}
    & \mathbf{x}_l = \left[ x_{l, l_{\text{key}}} \right]^T, 1 \le l_{\text{key}} \le N_{L_{\text{key}}}   \label{eq:obs_vec}
\end{align}
\begin{align}
    & x_{l, l_{\text{key}}} = \max \left\{ \frac{ \sum\limits_{t=1}^{N_T} \sum\limits_{\tau=0}^{N_\tau-1} \left| \frac{ F_{l_{\text{key}}}^{<l>} }{ F_{l_{\text{key}}} } \right| }{N_T N_\tau} ,1\right\} \label{eq:obs_single} \\
    & F_{l_{\text{key}}}^{<l>} = \sum\limits_{i = 1}^{{N_G}} {T_{l_{\text{key}},J(i)}^{<l>}\widehat {P}_{i,(t,\tau )}^{<l>}}  - \sum\limits_{j = 1}^{{N_D}} {T_{l_{\text{key}},j}^{<l>}D_{j,t + \tau }} \label{eq:obs_pfl}\\
    & F_{l_{\text{key}}} = \sum\limits_{i = 1}^{{N_G}} {T_{l_{\text{key}},J(i)}\widehat {P}_{i,(t,\tau )}}  - \sum\limits_{j = 1}^{{N_D}} {T_{l_{\text{key}},j} D_{j,t + \tau }} \label{eq:obs_pf}
\end{align}

The observation vector \eqref{eq:obs_vec} is designed with $N_{L_{\text{key}}}$ dimensions, each representing one key transmission line $l_{\text{key}}$ being observed. The corresponding element \eqref{eq:obs_single} denotes the average change rate of the power flow on line $l_{\text{key}}$ in all look-ahead windows before and after the outage of line $l$, where $N_T$ and $N_\tau$ are the total number of time slots and the duration of a look-ahead window, respectively. The average change rate is ignored when below 1, as it becomes less challenging in terms of power flow limits. $F_{l_{\text{key}}}^{<l>}$ \eqref{eq:obs_pfl} and $F_{l_{\text{key}}}$ \eqref{eq:obs_pf} represent the power flow on line $l_{\text{key}}$ with and without an outage in line $l$, where $N_G$ and $N_D$ are the total number of thermal units and buses. $T_{l_{\text{key}},j}^{<l>}$ and $T_{l_{\text{key}},j}$ denote the power transfer distribution factors, and function $J(i)$ represents the bus number of thermal unit $i$. $D_{j,t + \tau }$ denotes the net demand of bus $j$ at time $t+\tau$. $\widehat {P}_{i,(t,\tau )}^{<l>}$ and $\widehat {P}_{i,(t,\tau )}$ represent the real power output of thermal unit $i$ at time $t+\tau$, which can be obtained from historical data or by solving the DC optimal power flow (DCOPF) problem in the following section.

With the observation vectors for all the N-1 cases, a variety of clustering algorithms can be applied to aggregate them using k-means, hierarchical clustering, and so on \cite{xuSurveyClusteringAlgorithms2005}. After the clustering step, the cluster where the original N-1 case of the agent belongs becomes the network scenario set for further evaluation, and the number of cases it contains is defined as the total number of network scenarios $S_T$. Combined with $S_D$ demand scenarios generated earlier, the $S_T S_D$ scenarios for performance evaluation of the agent are therefore generated.

\section{Multi-scenario Performance Evaluation}

\subsection{Baseline Generation}

In order to properly evaluate the performance of the agent, a baseline for comparison is necessary. In this paper, the baseline of dispatch decisions is generated by solving a DCOPF problem. In practice, such baseline can also be obtained from historical data or the experience of system operators.

The DCOPF model for look-ahead economic dispatch is formulated as follows:
\begin{equation}
    \mathop {\min }\limits_{\widehat{\bf{P}}_{(t,\tau)}^{({s_T},{s_D})}} \sum\limits_{\tau = 0}^{N_\tau - 1} C(\widehat{\bf{P}}_{(t,\tau )}^{({s_T},{s_D})}) ,\quad \forall t,\forall ({s_T},{s_D})  \label{eq:dcopf_obj}
\end{equation}
subject to:
\begin{align}
    & \sum\limits_{i = 1}^{{N_G}} {\widehat {P}_{i,(t,\tau )}^{({s_T},{s_D})}} = \sum\limits_{j = 1}^{{N_D}} {D_{j,t + \tau }^{({s_D})}} ,\quad \forall t,\forall \tau ,\forall ({s_T},{s_D})   \label{eq:dcopf_cons1}
\end{align}
\begin{align}
    & \underline {{P_i}} \le \widehat {P}_{i,(t,\tau )}^{({s_T},{s_D})} \le \overline {{P_i}} ,\quad \forall i,\forall t,\forall \tau ,\forall ({s_T},{s_D})   \label{eq:dcopf_cons2}\\
    & - R{D_i} \le \widehat {P}_{i,(t,\tau )}^{({s_T},{s_D})} - \widehat {P}_{i,(t,\tau  - 1)}^{({s_T},{s_D})} \le R{U_i},   \nonumber   \\
    & \quad \quad \quad \forall i,\forall t,\forall \tau ,\forall ({s_T},{s_D})   \label{eq:dcopf_cons3}\\
    & - \overline{F_l} \le \sum\limits_{i = 1}^{{N_G}} {T_{l,J(i)}^{({s_T})}\widehat {P}_{i,(t,\tau )}^{({s_T},{s_D})}}  - \sum\limits_{j = 1}^{{N_D}} {T_{l,j}^{({s_T})}D_{j,t + \tau }^{({s_D})}}  \le \overline{F_l},    \nonumber   \\
    & \quad \quad \quad \forall 1 \le l \le {N_L},\forall t,\forall \tau ,\forall ({s_T},{s_D})   \label{eq:dcopf_cons4}
\end{align}
where $\widehat{\bf{P}}_{(t,\tau)}^{({s_T},{s_D})} = [\widehat {P}_{i,(t,\tau )}^{({s_T},{s_D})}], i=1,2,\dots,N_G$ denotes a vector of real power output baseline of all the thermal units at time $t+\tau$ in the look-ahead window $[t, t+N_\tau-1]$ in scenario $(s_T, s_D)$. Function $C(\cdot)$ defines the total operation cost of all the thermal units. $D_{j,t + \tau }^{({s_D})}$ denotes the net demand of bus $j$ at time $t+\tau$ in demand scenario $s_D$. $\overline {{P_i}}$ and $\underline {{P_i}}$ denote the upper and lower limits of power output, and $RU_i$ and $RD_i$ denote the ramping limits of thermal unit $i$, respectively. $\overline{F_l}$ represents the power flow limit of transmission line $l$, and $N_L$ is the number of transmission lines in total. $T_{l,j}^{({s_T})}$ denotes the power transfer distribution factor in network scenario $s_T$. Unless noted otherwise, all $\forall t$ in this paper indicates $\forall t=1,2,\dots,N_T$, representing all time slots. All $\forall \tau$ indicates $\forall \tau=0,1,\dots,N_\tau-1$, representing the entire look-ahead window. All $\forall i$ indicates $\forall i=1,2,\dots,N_G$, representing all thermal units. The symbol $\forall ({s_T},{s_D})$ indicates all combined scenarios of the network and demand, i.e., $\forall s_T=1,2,\dots,S_T,\;\forall s_D=1,2,\dots,S_D$.

The objective \eqref{eq:dcopf_obj} of the DCOPF model is to minimize the total operation cost in each look-ahead window in all scenarios with respect to the power output of thermal units. The constraints consist of power balance constraints \eqref{eq:dcopf_cons1}, power output constraints of thermal units \eqref{eq:dcopf_cons2}, ramping rate constraints of thermal units \eqref{eq:dcopf_cons3} and power flow constraints \eqref{eq:dcopf_cons4}. The DCOPF model above is a transformation of the RL framework of the agent. Therefore, the baseline of power output of thermal units can be obtained by solving the model above in all the scenarios.

\subsection{Evaluation Metrics}

The multi-scenario evaluation metrics are divided into economy and security metrics, and all the metrics are integrated from the simulation results of the individual scenarios. The evaluation metric for economy is relative cost error (RCE), defined as follows:
\begin{align}
    & RCE = \frac{ \sum\limits_{s_T = 1}^{S_T} \sum\limits_{s_D = 1}^{S_D} RCE^{(s_T,s_D)} }{ S_T S_D } \\
    & RCE^{(s_T,s_D)} = \frac{ \sum\limits_{t = 1}^{N_T} \sum\limits_{\tau  = 0}^{N_\tau - 1} C({\bf{P}}_{(t,\tau )}^{(s_T,s_D)}) }{\sum\limits_{t = 1}^{N_T} \sum\limits_{\tau  = 0}^{N_\tau - 1} C(\widehat {\bf{P}}_{(t,\tau )}^{(s_T,s_D)}) } - 1,\quad \forall (s_T,s_D)
\end{align}
where ${\bf{P}}_{(t,\tau)}^{({s_T},{s_D})} = [P_{i,(t,\tau )}^{({s_T},{s_D})}], i=1,2,\dots,N_G$ denotes a vector of real power output of all the thermal units produced by the agent in scenario $(s_T, s_D)$. The RCE of the dispatch decisions generated by the agent is defined as the relative mean error of total operation cost in all look-ahead windows in each scenario with respect to the baseline.

The evaluation metrics for security include total relative violation of constraints (RVS), maximum relative violation of constraints (RVM), average number of violated constraints (NVC), average number of time slots with violated constraints (NVT) and availability rate ($\eta$):
\begin{align}
    & RVS = \sum\limits_{{s_T} = 1}^{{S_T}} {\sum\limits_{{s_D} = 1}^{{S_D}} {RV{S^{({s_T},{s_D})}}} }  \label{eq:metric_RVS}\\
    & RVM = \mathop {\max }\limits_{({s_T},{s_D})} RV{M^{({s_T},{s_D})}}    \label{eq:metric_RVM}  \\
    & NVC = \frac{{\sum\limits_{{s_T} = 1}^{{S_T}} {\sum\limits_{{s_D} = 1}^{{S_D}} {NV{C^{({s_T},{s_D})}}} } }}{{{S_T}{S_D}}}  \label{eq:metric_NVC} \\
    & NVT = \frac{{\sum\limits_{{s_T} = 1}^{{S_T}} {\sum\limits_{{s_D} = 1}^{{S_D}} {NV{T^{({s_T},{s_D})}}} } }}{{{S_T}{S_D}}}  \label{eq:metric_NVT} \\
    & \eta  = \frac{{{S_T}{S_D} - \sum\limits_{{s_T} = 1}^{{S_T}} {\sum\limits_{{s_D} = 1}^{{S_D}} {{\mathop{\rm sgn}} \left( {NV{T^{({s_T},{s_D})}}} \right)} } }}{{{S_T}{S_D}}} \times 100\%   \label{eq:metric_eta}
\end{align}

RVS \eqref{eq:metric_RVS} and RVM \eqref{eq:metric_RVM} are two metrics characterizing the extent of violation of constraints in general. The RVS of the dispatch decisions is defined as the summation of all scenarios, and the RVM is defined as the maximum value. In each scenario, the RVS is defined as the total relative violation of power output constraints, ramping rate constraints and power flow constraints in all look-ahead windows, and the RVM is the corresponding maximum value. $VO_{i,(t,\tau )}^{({s_T},{s_D})}$, $VR_{i,(t,\tau )}^{({s_T},{s_D})}$ and $VL_{l,(t,\tau )}^{({s_T},{s_D})}$ are relative violation values of the three aforementioned types of constraints, and the limits of the constraints are utilized to measure the level of violation:
\begin{align}
    & RV{S^{({s_T},{s_D})}} = \sum\limits_{t = 1}^{{N_T}} \sum\limits_{\tau  = 0}^{{N_\tau } - 1}
        \left( \begin{array}{l}
            \sum\limits_{i=1}^{N_G} VO_{i,(t,\tau )}^{({s_T},{s_D})} +  \\
            \sum\limits_{i=1}^{N_G} VR_{i,(t,\tau )}^{({s_T},{s_D})} +  \\
            \sum\limits_{l=1}^{N_L} VL_{l,(t,\tau )}^{({s_T},{s_D})}
        \end{array} \right) \\
    & RV{M^{({s_T},{s_D})}} = \mathop {\max }\limits_{\scriptstyle1 \le t \le {N_T}\atop \scriptstyle0 \le \tau  \le {N_\tau } - 1}
        \left\{ \begin{array}{l}
            \mathop {\max }\limits_{1 \le i \le N_G} \{VO_{i,(t,\tau )}^{({s_T},{s_D})}\},  \\
            \mathop {\max }\limits_{1 \le i \le N_G} \{VR_{i,(t,\tau )}^{({s_T},{s_D})}\},  \\
            \mathop {\max }\limits_{1 \le l \le N_L} \{VL_{l,(t,\tau )}^{({s_T},{s_D})}\}
        \end{array} \right\}    \\
    & VO_{i,(t,\tau )}^{({s_T},{s_D})} = \max \left\{ \begin{array}{l}
        \frac{{P_{i,(t,\tau )}^{({s_T},{s_D})} - \overline {{P_i}} }}{{\overline {{P_i}} }},    \\
        \frac{{\underline {{P_i}}  - P_{i,(t,\tau )}^{({s_T},{s_D})}}}{{\underline {{P_i}} }},    \\
        0
        \end{array} \right\}
\end{align}
\begin{align}
    & VR_{i,(t,\tau )}^{({s_T},{s_D})} = \max \left\{ \begin{array}{l}
        \frac{{P_{i,(t,\tau )}^{({s_T},{s_D})} - P_{i,(t,\tau  - 1)}^{({s_T},{s_D})} - R{U_i}}}{{R{U_i}}},\\
        \frac{{P_{i,(t,\tau  - 1)}^{({s_T},{s_D})} - P_{i,(t,\tau )}^{({s_T},{s_D})} - R{D_i}}}{{R{D_i}}},\\
        0
        \end{array} \right\} \\
    & VL_{l,(t,\tau )}^{({s_T},{s_D})} = \max \left\{ \begin{array}{l}
        \frac{\left| \begin{array}{l}
            \sum\limits_{i = 1}^{{N_G}} {T_{l,J(i)}^{({s_T})}P_{i,(t,\tau )}^{({s_T},{s_D})}} -    \\
            \sum\limits_{j = 1}^{{N_D}} {T_{l,j}^{({s_T})}D_{j,t + \tau }^{({s_D})} }
            \end{array} \right|}{\overline{F_l}} - 1,\\
        0
        \end{array} \right\}
\end{align}

NVC \eqref{eq:metric_NVC}, NVT \eqref{eq:metric_NVT} and availability rate $\eta$ \eqref{eq:metric_eta} are metrics characterizing the proportion of unusable dispatch decisions, regardless of the extent of violation. The NVC and NVT metrics for dispatch decisions are derived by averaging over all scenarios. In each scenario, the NVC is defined as the total number of violated constraints in all look-ahead windows, and the NVT is the total number of time slots with violated constraints. Availability rate $\eta$ is defined as the proportion of scenarios without any violation of constraints.
\begin{align}
    & NV{C^{({s_T},{s_D})}} = \frac{{\sum\limits_{t = 1}^{{N_T}} {\sum\limits_{\tau  = 0}^{{N_\tau } - 1} {\left( \begin{array}{l}
        \sum\limits_{i = 1}^{{N_G}} \mathop{\rm sgn} VO_{i,(t,\tau )}^{({s_T},{s_D})} + \\
        \sum\limits_{i = 1}^{{N_G}} \mathop{\rm sgn} VR_{i,(t,\tau )}^{({s_T},{s_D})} + \\
        \sum\limits_{l = 1}^{{N_L}} \mathop{\rm sgn} VL_{i,(t,\tau )}^{({s_T},{s_D})}
        \end{array} \right)} } }}{{{N_T}{N_\tau }(2{N_G} + {N_L})}} \\
    & NV{T^{({s_T},{s_D})}} = \frac{{\sum\limits_{t = 1}^{{N_T}} {\sum\limits_{\tau  = 0}^{{N_\tau } - 1} {{\mathop{\rm sgn}} \left( \begin{array}{l}
        \sum\limits_{i = 1}^{{N_G}} VO_{i,(t,\tau )}^{({s_T},{s_D})} + \\
        \sum\limits_{i = 1}^{{N_G}} VR_{i,(t,\tau )}^{({s_T},{s_D})} + \\
        \sum\limits_{l = 1}^{{N_L}} VL_{i,(t,\tau )}^{({s_T},{s_D})}
        \end{array} \right)} } }}{{{N_T}{N_\tau }}}
\end{align}

The above metrics constitute a comprehensive evaluation framework for the agent. By calculating the value of the metrics after obtaining the simulation results of the agent in all scenarios, the performance of the agent can be properly evaluated.

\section{Case Study}

The evaluation of an RL agent is conducted in a modified IEEE 30-bus power system. This agent is designed for the look-ahead economic dispatch and is tested for a total of 40 days ($N_T=3840$). The interval of time slots is set as 15~min, and the length of the look-ahead window $N_\tau = 16$.

\subsection{Scenarios}

To fully evaluate agent performance, 123 scenarios are designed by combining $S_T=3$ different network scenarios with $S_D=41$ demand scenarios. As mentioned above, the demand scenarios are generated by randomly shifting the original demand level used to train the agent. To be specific, the original demand curve is multiplied with 41 coefficients from 80\% to 120\% with a constant interval of 1\%, each added with normally distributed random factors for all time slots.

\begin{figure}
    \centering
    \includegraphics[width=0.5\textwidth]{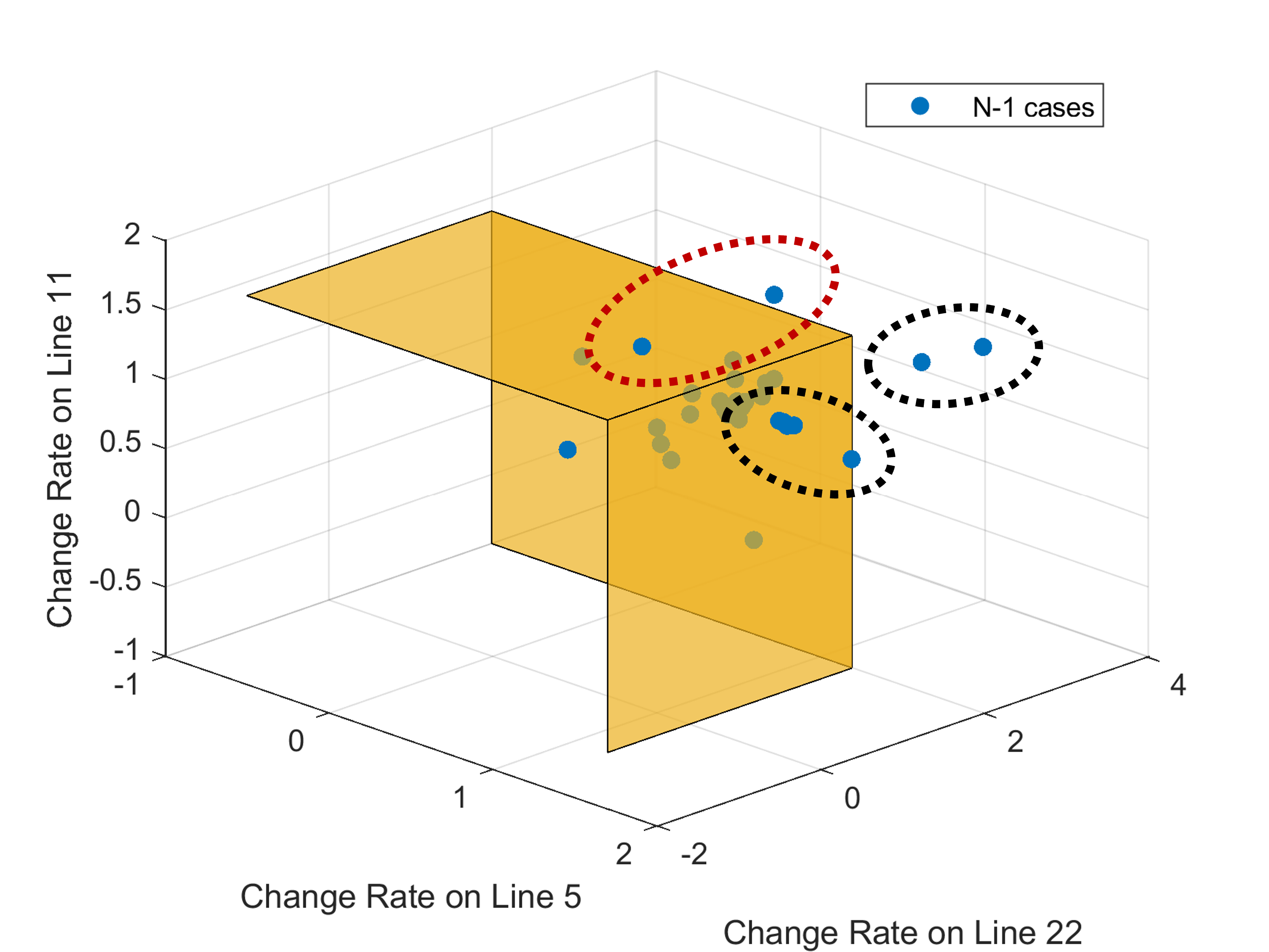}
    \caption{Network clustering results of all N-1 cases.}
    \label{fig:cluster}
\end{figure}

The network scenarios, on the other hand, are generated with network clustering. For all 41 N-1 cases, the average change rates of power flow on key transmission lines 5, 11 and 22 are depicted in Fig~\ref{fig:cluster}. Here, the N-1 cases are aggregated into 4 clusters. The first cluster (inside the semi-cube area) represents the cases whose power flow changes little or becomes less challenging, while others (each represented with an ellipse) correspond to significant increase in the power flow on one key transmission line, indicating greater challenge on power flow limits. As the agent in this section is designed for the N-1 case where line 10 becomes unavailable, the cluster it belongs to (the red ellipse) becomes the network scenario set, which contains N-1 cases of lines 10, 13, and 14.

\subsection{Evaluation Results}

\begin{figure}
    \centering
    \includegraphics[width=0.5\textwidth]{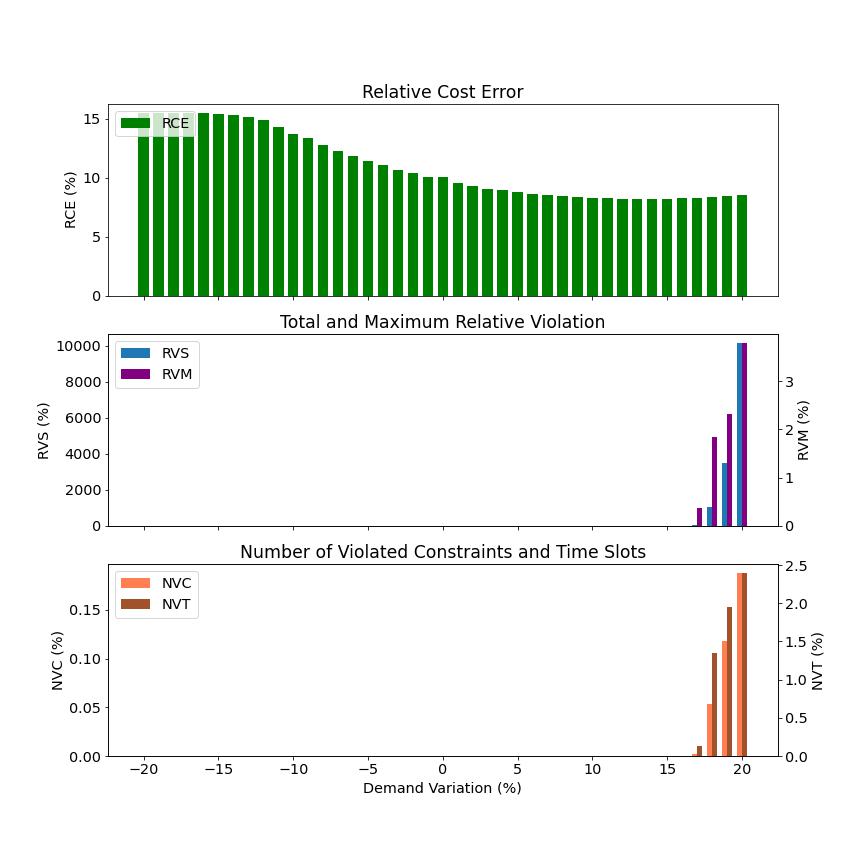}
    \caption{Evaluation results of the agent in a typical network scenario.}
    \label{fig:eval}
\end{figure}

Based on the scenarios generated, the performance of the agent is evaluated using the metrics proposed in this paper. The evaluation results are listed in Table~\ref{tab:eval1}, and the detailed results in one network scenario are depicted in Fig~\ref{fig:eval}, where line 13 becomes unavailable instead of the original line 10.

\begin{table}
    \centering
    \caption{Multi-Scenario Evaluation Results of the Agent}    \label{tab:eval1}
    \begin{tabular}{llllll}
        \toprule
        RCE(\%) & RVS(\%) & RVM(\%) & NVC(\%) & NVT(\%) & $\eta$(\%)   \\
        \midrule
        11.0 & 545763.1 & 13.2 & 0.035 & 0.33 & 84.6    \\
        \bottomrule
    \end{tabular}
\end{table}

The evaluation results indicate that the agent is capable of adapting to a certain level of environment disturbance. In Fig.~\ref{fig:eval}, with demand varying between $-20\%$ and $20\%$, the agent produces feasible dispatch decisions in 37 out of 41 scenarios. On one hand, when demand is too high, the dispatch decisions of the agent violate the power flow constraints in a few look-ahead windows, which becomes more severe in terms of both the proportion of unusable decisions and the extent of violation as demand increases. On the other hand, when demand becomes too low, although there is no violation of constraints, the operation cost of the dispatch decisions increases gradually with respect to the baseline, suggesting decreased effectiveness. In addition, while the violation in some extreme scenarios (near 20\% demand variation) is significant, the majority remains acceptable, which reflects in Table~\ref{tab:eval1} as well. In the table, despite the high RVS and RVM values suggesting dramatic violation in certain scenarios, the NVC and NVT values are relatively low, suggesting general usability. Judging from all the metric values, the evaluated agent possesses the basic ability of adapting to different scenarios, while under some circumstances it loses such ability and needs further improvement.

\subsection{Comparison Between Different Agents}

The evaluation approach is further employed on several other RL agents apart from the previous one, and the results are listed in Table~\ref{tab:eval2}. The metric values in the table are all percentage values, same as Table~\ref{tab:eval1}. In the first column of the table, the names of the agents suggest the number of episodes in their training process, and RL-50 correspond to the one used in the previous section.

\begin{table}
    \centering
    \caption{Evaluation Results of Different Agents}    \label{tab:eval2}
    \begin{tabular}{lllllll}
        \toprule
        Agent Name & RCE & RVS & RVM & NVC & NVT & $\eta$   \\
        \midrule
        RL-50 & 11.0 & 545763.1 & 13.2 & 0.035 & 0.33 & 84.6     \\
        RL-100 & 11.2 & 505855.3 & 12.9 & 0.033 & 0.31 & 82.9    \\
        RL-150 & 11.3 & 587842.5 & 12.7 & 0.038 & 0.33 & 81.3    \\
        \bottomrule
    \end{tabular}
\end{table}

Here, the evaluation results of the agents provide a reference for the selection of training episodes. In Table~\ref{tab:eval2}, with the increase of training episodes of the agents, the evaluation results become slightly different. For RVS, RVM, NVC and NVT, the lowest values (best situation) occur when RL-100 is selected. The dispatch decisions of both RL-50 and RL-150 are less secure in terms of those metrics. This phenomenon is probably due to the fact that more episodes can improve the ability to capture the characteristics of the training set, while increasing the risk of over-fitting at the same time. In addition, judging from RCE and $\eta$, the further training process compromises the accuracy and economy of the dispatch decisions of the agent. In summary, it can be drawn from the evaluation results that 100 episodes make a relatively appropriate choice for training in this case.

\section{Conclusion}

This paper proposes an evaluation approach for look-ahead economic dispatch in order to properly analyze the performance of RL agents under multiple scenarios. A network clustering method is developed to generate the scenarios, and a series of evaluation metrics are designed for each scenario considering both economy and security. Evaluation results of multiple agents for a modified IEEE 30-bus system show that the proposed approach can effectively assess the adaptability of an agent. In addition, this approach can be utilized to offer suggestions for the value of key parameters in the RL algorithm by comparing results among different agents. In brief, this work will contribute to the improvement of RL algorithms in look-ahead dispatch, therefore increasing the intelligence of power system operation.


\bibliographystyle{IEEEtran}
\bibliography{Reference}

\end{document}